\documentclass[amsmath,amssymb,twocolumn]{revtex4}
\usepackage[parfill]{parskip}    
\usepackage{graphicx}
\usepackage{amssymb}
\usepackage{epstopdf}
\usepackage{color}
\usepackage{graphicx}
\usepackage{pdfpages}

\begin{document}
\title{Scale--dependent Hayward black hole and the generalized uncertainty principle}
\author{E. Contreras $^1$\footnote{On leave from Universidad Central de Venezuela}\footnote{ejcontre@espol.edu.ec}
and P. Bargue\~no $^2$\footnote{p.bargueno@uniandes.edu.co}}
\affiliation{
$^1$ Escuela Superior Polit\'ecnica del Litoral, ESPOL, Facultad de Ciencias Naturales y Matem\'aticas, Apartado Postal 09-01-5863, Campus Gustavo Galindo Km 30.5 V\'ia Perimetral, Guayaquil, Ecuador\\
$^2$Departamento de F\'{\i}sica, Universidad de los Andes, Cra.1E
No.18A-10, Bogot\'a, Colombia \\
}
\begin{abstract}
In this work we present a technique to obtain bounds on the generalized uncertainty principle deformation parameter by
using an improved Schwarzschild solution represented by the Hayward metric in the context of scale--dependent gravity.
Specifically, this deformation parameter can be interpreted in terms of a running parameter which controls the deviation from
the standard Einstein--Hilbert action in the scale--dependent scenario.
\end{abstract}

\maketitle

\section{Introduction}
\label{intro}

The interior of black holes (BHs) is plagued with uncertainties due to the presence of 
singularities, predicted some time ago \cite{Hawking}. Until a complete theory for quantum gravity is developed,
these singularities are expected to be smoothed out by some effective theory operating, presumably, at the Planck scale. 
In this sense, the study of regular ({\it i. e.}, non--singular) BH solutions, is well motivated.
Among all the regular models for BHs considered in the literature (for an incomplete list see, for example,
\cite{Sakharov1966,Gliner1966,Bardeen1968,Dymnikova1992,Mars1996,Ayon1998,Bronnikov2000,hayward2005,vagenas2014}),
particular interest has raised recently regarding the Hayward solution in the context of the so--called Planck stars
\cite{Barrau2014,Rovelli2014,DeLorenzo2015,Christodoulou2016,Rovelli2017}, which are hypothetical objects formed at the end of 
gravitational collapse, which stops before the central singularity 
is formed, yielding to the formation of a central core. Planck stars, which can be much larger than the Planck length, 
are conjectured to be possible sources of radio and gamma--ray bursts \cite{Rovelli2017}. Therefore, the interest in the study 
of these kind of objects comes from both the theoretical and the experimental side because their represent a
speculative but realistic possibility to observe quantum gravity effects \cite{Rovelli2017}.

Regarding these effects, different ways to introduce such a corrections to the well 
known classical BH solutions have been considered by promoting the metric function to a scale dependent one, which is known from the 
renormalization
group flow (see  \cite{Bonanno:2000ep,Bonanno:2006eu,Koch:2013owa,Koch:2014cqa} and references therein). 
A different approach, based in an effective action, consists in promoting the coupling constants of the theory to fields 
on the underlying space--time. This procedure leads to modified Einstein field equations to be solved for both the metric
tensor and the scale--dependent couplings (fields) involved.
This idea was previously investigated by Weinberg thought the well know Weinberg's Asymptotic
Safety program 
\cite{
Weinberg:1979,
Wetterich:1992yh,
Dou:1997fg,
Souma:1999at,
Reuter:2001ag,
Fischer:2006fz,
Percacci:2007sz,
Litim:2008tt}. The same approach was extended in Ref. \cite{cai2010} to study
 study black hole solutions in an asymptotically safe gravity theory including higher
derivative terms and running gravitational couplings.
Recently, the technique has been employed in both three--dimensional and 
\cite{koch2016,Rincon:2017ypd,Rincon:2017goj,Rincon:2017ayr,Rincon2018,Rincon2018b} 
four--dimensional BHs \cite{koch2015a, koch2015,koch2015b, contreras2018, contreras2018b, Contreras2018, rincon2018} and
in a cosmological context \cite{Hernandez2018}.
 
Another technique is related to the Generalized Uncertainty Principle (GUP) which consists in a modification of the
Heisenberg uncertainty principle as a consequence of the existence of a minimum length when gravitational effects are
switched on \cite{Mead1964,Amati1989,Maggiore1993,Maggiore1994,Maggiore1993b,
Garay1995,Scardigli1999,Hossenfelder2003,Bambi2008,Hossain2010,Kempf1995,Kempf1997,
Brau1999,Magueijo2002,Magueijo2003,vagenas2004,Magueijo2005,Cortes2005,Ghosh2007,Das2008,
Das2009,Ali2009,Das2010,Ali2011,bargueno2015}. More rencently, 
the GUP parameter have been computed by conjecturing that the GUP--deformed BH temperature of a Schwarzschild BH and the modified 
Hawking temperature of a quantum--corrected Schwarzschild BH are the same \cite{scardigli2015,Elias2017,vagenas2018,scardigli2018}. From the
experimental point of view, the feasibility of detecting these Planck--scale effects by measuring the
deformation parameter at the laboratory has recently attracted considerable attention \cite{Brucker2012,Marin2013,Bawaj2015,
Plenio2017}.

In this work we interpret the Hayward BH solution as a modification of the 
Schwarzschild BH which arises as a consequence of the scale--dependence of the couplings on the underlying theory 
\cite{koch2015,koch2015b,koch2016,Rincon:2017ypd,Rincon:2017goj}. In this sense, the BH temperature can be 
considered as a scale--dependent corrected quantity, which we expect to coincide with corrections coming from
other approaches. Specifically, the scale--dependent treatment should give a corrected temperature consistent with
the corrections coming from the GUP. This procedure will allow us to obtain some bounds on the two parameters characterizing
the approaches, named the GUP and running parameters.

The work is organized as follows. Section \ref{sdg} is devoted to summarize the main aspects of 
scale--dependent gravity. In Sect. \ref{gup} we introduce a particular GUP, together with its corrected BH temperature. 
The scale--dependent Hayward solution is studied in Sect. \ref{Hayward}. Sec. \ref{comparison} is devoted to a comparison between
scale--dependent and GUP corrections to the BH temperature, and final comments are left to the concluding remarks on 
Sect. \ref{remarks}

\section{Scale--dependent gravity}
\label{sdg}

The effective Einstein--Hilbert here considered reads
\begin{eqnarray}\label{action}
S[g_{\mu\nu},k]=\int \mathrm{d}^{4}x\sqrt{-g}\frac{R}{16 \pi G_{k}}+S_{matter},
\end{eqnarray}
with $G_{k}$ standing for the scale--dependent gravitational coupling, and $S_{matter}$ is the action for the matter sector.

After variations with respect to the metric field, $g_{\mu\nu}$, the modified Einstein's field equations can be written as
\begin{eqnarray}\label{einstein}
G_{\mu\nu} = 8\pi G_{k} T^{eff}_{\mu\nu},
\end{eqnarray}
where
$T^{eff}_{\mu\nu}$ is the effective energy momentum tensor defined as
\begin{eqnarray}\label{eff}
T^{eff}_{\mu\nu}:=T_{\mu\nu}-\frac{1}{ 8\pi G_{k}}\Delta t_{\mu\nu},
\end{eqnarray}
$T_{\mu\nu}$ corresponds to the matter energy--momentum tensor and $\Delta t_{\mu\nu}$, given by
\begin{eqnarray}\label{nme}
\Delta t_{\mu\nu}=G_{k}\left(g_{\mu\nu}\square -\nabla_{\mu}\nabla_{\nu}\right)G_{k}^{-1},
\end{eqnarray}
is the so--called non--matter energy--momentum tensor.

Variations  with respect to the scale--field, $k(x)$, lead to
\begin{eqnarray}\label{scale}
\frac{\delta S[g_{\mu\nu},k]}{\delta k}=0.
\end{eqnarray}
In order to avoid the use of beta functions, and following \cite{koch2015,koch2015b,koch2016, contreras2018},
all the couplings present in the action are promoted to fields depending on space--time coordinates. In our case, 
this translates into $G_{0}\rightarrow G(x)$, where $G_{0}$ is the classical ({\it i. e.} non--running) gravitational constant.

For the purpose of the present work, we consider a static and spherically symmetric space--time with a line element parametrized 
as
\begin{eqnarray}\label{metricp}
ds^2=-f(r)dt^2 + f(r)^{-1}dr^2+r^2 d\Omega^2. 
\end{eqnarray}

As in a previous work \cite{contreras2018}, after replacing Eq. (\ref{metricp}) in Eq. (\ref{einstein}), three independent 
differential equations for the four independent fields $f(r)$, $G(r)$, $T^{0}_{0}$ and $T^{2}_{2}$, are obtained and, 
in order to decrease the number of degrees of freedom, the null energy condition (NEC) is demanded 
\cite{koch2015b,koch2016,contreras2018}. With the parametrization of Eq. (\ref{metricp}), both $G_{\mu\nu}$ and $T_{\mu\nu}$ 
saturate the NEC and, therefore, $\Delta t_{\mu\nu}$ must saturate it also for consistency. This last condition on 
$\Delta t_{\mu\nu}$ leads to
\begin{eqnarray}\label{gr}
G(r)=\frac{G_{0}}{1+\epsilon r},
\end{eqnarray}
where $\epsilon\ge 0$ is a parameter with 
dimensions of inverse of length. At this point a couple of comments are in order. First, 
it is worth mentioning that, in the limit $\epsilon\to 0$, $G(r)=G_{0}$, $\Delta t_{\mu\nu}=0$ 
and the classical Einstein's field equations are recovered. For this reason, $\epsilon$ is called the running parameter, which controls the strength of the scale--dependence \cite{koch2016,Rincon:2017ypd,Rincon:2017goj,Rincon:2017ayr,contreras2018}. Second, the Newton coupling $G(r)$ we have obtained should be thought as an effective field which takes into account the deviations of the theory in the scale--dependence scenario from General Relativity. And third, this $G(r)$ is not at the same level of that
considered in asymptotic safety \cite{
Weinberg:1979,
Wetterich:1992yh,
Dou:1997fg,
Souma:1999at,
Reuter:2001ag,
Fischer:2006fz,
Percacci:2007sz,
Litim:2008tt} because we don not have an explicit relation between the energy
scale and the $r$ coordinate. 

\section{GUP and BH temperature}
\label{gup}

Among the various forms that have been suggested in the past of deformations of the commutation relations, 
in this work we are interested in the particular case given by \footnote{We shall work with $\hbar=c=1$ but explicitly show
the Newton constant, $G_{0}$. With this choice, the Planck mass is $m_{p}=G_{0}^{-1/2}$.}
\begin{equation}
[\hat x,\hat p] = \mathrm{i} \left(1+\beta\left(\frac{\hat p}{m_{p}}\right)^2 \right)
\end{equation}
which is equivalent, for states satisfying $\langle \hat p\rangle^2=0$, to a modification of the uncertainty principle which
can be expressed as
\begin{equation}
\Delta x \Delta p\ge 
\frac{1}{2}\left(1+\beta\left(\frac{\Delta p}{m_{p}}\right)^2 \right)
\end{equation}
where $\beta$ is the GUP parameter.

The effect of this GUP on the standard Hawking temperature for a Schwarzschild BH follows from the
argument of the Heisenberg microscope \cite{Heisenberg1927} together with the consideration of an ensemble of unpolarized photons (of Hawking
radiation) outside the horizon with uncertainty given by the GUP, together with their energy equipartition
\cite{Scardigli1995,Adler2001,Cavaglia2004,Cavaglia2003,Susskind2005,Nouicer2007}
. 
After inverting the mass--temperature relation predicted by the GUP (see, for example, Ref. \cite{Elias2017}) we arrive to
\begin{equation}
T_{\mathrm{GUP}}=\frac{\pi}{\beta}\left(M_{0}-\sqrt{M_{0}^2-\frac{\beta}{\pi^2}m_{P}^2} \right),
\end{equation}
where $M_{0}$ is the BH mass. As the GUP parameter $\beta$ is expected to be or the order of unity, for BHs with 
$M_{0}\gg m_{P}$ we can
expand in powers to $\beta$ to get
\begin{equation}
T_{\mathrm{GUP}}=\frac{1}{8\pi G M_{0}}\left(1+\frac{\beta m_{P}^2}{4\pi^2 M_{0}^2}+... \right),
\end{equation}
recovering the usual Hawking spectrum at zero order in $\beta$.

We stress that this shifted Hawking temperature is due to the thermal character of the GUP corrections 
which can be also computed, for example, from a canonical point of view \cite{Bargueno2015}. 
In the next sections we will show that this
correction to the Hawking temperature can be associated with a running mechanism developed for the Hayward BH.

\section{Scale--dependent Hayward solution}
\label{Hayward}
In this section, we reformulate the Hayward solution \cite{hayward2005} as a modified Schwarzschild solution in the context 
of scale--dependent gravity. 
More precisely, given the Hayward solution \cite{hayward2005}
\begin{eqnarray}\label{hayward}
f(r)=1-\frac{2 G_{0} M_{0} r^{2}}{r^{3}+2L_{0}^{2}M_{0}}, 
\end{eqnarray}
where $M_{0}$ is the BH mass, $L_{0}$ corresponds to the Hubble length, and following the idea presented in a previous work 
\cite{contreras2018}, the scale--dependent solution can be obtained simply 
by replacing 
$2L_{0}^{2}M_{0}\to G_0^3 M_0 \epsilon ^2$ to obtain
\begin{eqnarray}\label{sdhay}
f(r)=1-\frac{2 G_0 M_0 r^2}{r^3+G_0^3 M_0 \epsilon ^2}.
\end{eqnarray}
Note that, as stated in \cite{contreras2018}, the deviation of Eq. (\ref{sdhay}) from 
the Schwarzschild solution
is now controlled by the running parameter $\epsilon$. Moreover, the Schwarzschild solution is recovered when $\epsilon\rightarrow0$.

The reformulation of Eq. (\ref{hayward}) in terms of scale--dependent gravity leads to a reinterpretation of the matter sector. 
In particular, the matter content $T_{\mu\nu}$ can be associated with certain anisotropic {\it vacuum} 
which appears in a regime where the scale dependence cannot be ignored \cite{contreras2018}. 
This can be shown by an explicit calculation which leads to
\begin{widetext}
\begin{eqnarray}\label{matterrun}
T^{0}_{0}=T^{1}_{1}&=&\frac{\epsilon  \left(-3 G_0 M_0 r^5+4 G_0^3 M_0 r^3 \epsilon ^2-6 G_0^4 M_0^2 r \epsilon  (2 r \epsilon +1)
+2 G_0^6 M_0^2 \epsilon ^4+2 r^6\right)}{8 \pi  G_0 r \left(G_0^3 M_0 \epsilon ^2+r^3\right){}^2}\\
T^{2}_{2}&=&T^{3}_{3}=\frac{\epsilon  \left(3 G_0^3 M_0 r^5 \epsilon ^2 \left(2 G_0 M_0+r\right)+12 G_0^4 M_0^2 r^4 \epsilon 
+3 G_0^6 M_0^2 r^2 \epsilon ^4 \left(r-4 G_0 M_0\right)\right)}{8 \pi  G_0 r \left(G_0^3 M_0 \epsilon ^2+r^3\right){}^3} \nonumber \\
&-&\frac{6 G_0^7 M_0^3 r \epsilon ^3+G_0^9 M_0^3 \epsilon ^6+r^9}{8 \pi  G_0 r \left(G_0^3 M_0 \epsilon ^2+r^3\right){}^3}.
\end{eqnarray}
\end{widetext}
Note that the non--matter energy momentum tensor, $\Delta t_{\mu\nu}$,
given by
\begin{widetext}
\begin{eqnarray}
\Delta t^{0}_{0}=\Delta t^{1}_{1}&=&\frac{r^5 \epsilon  \left(2 r-3 G_0 M_0\right)+2 G_0^3 M_0 r^2 \epsilon ^3 
\left(2 r-3 G_0 M_0\right)+2 G_0^6 M_0^2 \epsilon ^5}{r (r \epsilon +1) \left(G_0^3 M_0 \epsilon ^2+r^3\right){}^2}\\
\Delta t^{2}_{2}=\Delta t^{3}_{3}&=&\frac{2 G_0^3 M_0 r^2 \epsilon ^3 \left(r-3 G_0 M_0\right)+G_0^6 M_0^2 
\epsilon ^5+r^6 \epsilon }{r (r \epsilon +1) \left(G_0^3 M_0 \epsilon ^2+r^3\right){}^2}
\end{eqnarray}
\end{widetext}
acts as a counterterm which cancels the divergence of $T_{\mu\nu}$ at $r\to0$, in agreement with \cite{dymnikova2003}. Therefore,
we arrive to a regular effective energy momentum tensor whose components are expressed as
\begin{widetext}
\begin{eqnarray}
(T^{eff})^{0}_{0}=(T^{eff})^{1}_{1}&=&-\frac{3 G_0^3 M_0^2 \epsilon ^2 (r \epsilon +1)}{4 \pi  \left(G_0^3 M_0 \epsilon ^2+r^3\right){}^2}\\
(T^{eff})^{2}_{2}=(T^{eff})^{3}_{3}&=&-\frac{3 G_0^3 M_0^2 \epsilon ^2 (r \epsilon +1) \left(G_0^3 M_0 \epsilon ^2
-2 r^3\right)}{4 \pi  \left(G_0^3 M_0 \epsilon ^2+r^3\right){}^3}.
\end{eqnarray}
\end{widetext}
The effective density $-(T^{eff})^{0}_{0}$ is regular everywhere and decays faster that $r^{-3}$, as 
required in Ref. \cite{dymnikova2003}. Moreover, in the limit $r\rightarrow 0$, this $T^{eff}_{\mu\nu}$ becomes isotropic 
corresponding to a vacuum energy associated to an effective cosmological constant given by 
$\Lambda_{eff}=\frac{3}{4 \pi  G_0^3 \epsilon ^2}$. 
Note that the effective energy--momentum tensor vanishes when $\epsilon\rightarrow 0$, and the vacuum solution is
recovered in accordance with Birkhoff's theorem.

At this point, although BH thermodynamics could be studied as in previous 
works \cite{koch2015,koch2015b,koch2016,Rincon:2017ypd,Rincon:2017goj,Rincon:2017ayr}, in this work we will concentrate on
the BH temperature because this can be related to the corrected temperature obtained from the GUP approach, as it will be shown
in the next section.

\section{GUP and scale--dependent BH temperature}
\label{comparison}
The scale--dependent Hayward solution of Eq. (\ref{sdhay}) reads, for small $\epsilon$, as
\begin{eqnarray}\label{haywardserie}
f(r)=1-\frac{2 G_0 M_0}{r}+\frac{2 G_0^4 M_0^2 \epsilon ^2}{r^4}+\mathcal{O}(\epsilon^{3}).
\end{eqnarray}
Following Ref. \cite{Elias2017,vagenas2018}, for a line element parametrized as in Eq. (\ref{metricp}), the 
associated horizon temperature can be written as
\begin{eqnarray}\label{temperature}
T=\frac{1}{8\pi G_{0}M_{0}}(1+2\xi(a)+a\xi'(a)) ,
\end{eqnarray}
where $a=2G_{0}M_{0}$ corresponds to the Schwarzschild radius and the prime denotes derivative with respect to the
radial coordinate. Now, from Eq. (\ref{haywardserie}) it can be easily shown that 
\begin{eqnarray}\label{xi}
\xi(r)= \frac{2 G_0^4 M_0^2 \epsilon ^2}{r^4}.
\end{eqnarray}
On one hand, after replacing Eq. (\ref{xi}) in Eq. (\ref{temperature}), the scale--dependent--corrected temperature can 
be expressed by
\begin{eqnarray}
T_{\mathrm{SD}}\approx\frac{1}{8 \pi  G_0 M_0}+\frac{\epsilon ^2}{32 \pi  G_0 M_0^3}.
\end{eqnarray}
On the other hand, as stated previously, the temperature of a Schwarzschild BH is modified by the GUP as
\begin{eqnarray}
T_{\mathrm{GUP}}\approx\frac{1}{8 \pi  G_{0} M_{0}}+\frac{\beta  m_{p}^2}{32 \pi ^3 G_{0} M_{0}^3}.
\end{eqnarray}
Therefore, by comparing these expressions for the corrected BH temperature we get that the running and GUP parameters 
are related by
\begin{eqnarray}\label{epsi}
\beta = \left(\frac{\pi \epsilon}{m_{p}}\right)^2.
\end{eqnarray}

It is worth mentioning that, as commented in \cite{koch2015b}, the running parameter $\epsilon$ must be considered
small with respect to the other scales entering the problem such as $1/\sqrt{G_{0}}=m_{p}$.
More precisely, $\epsilon$ encodes the deviation from the classical solution and therefore its value is also experimentally expected to be very small in comparison with other integration constant with dimensions of energy.
 As a consequence, the GUP 
parameter $\beta$ is constrained by
\begin{equation}
\beta\le \pi^2.
\end{equation}

\section{Concluding remarks}
\label{remarks}
In this work we have obtained a new bound on the GUP deformation parameter, $\beta$, by means of 
a comparison between the corrected black hole temperature coming from the GUP and a scale--dependent approach. 
In order to do this,
we have shown that the Hayward regular black hole is a solution of an effective scale--dependent gravity theory, controlled
by certain running parameter, $\epsilon$. In this sense, scale--dependence provides a mechanism for the regularization of 
the Schwarzschild geometry. Moreover, as a bonus, we have shown that the simplest toy model describing Planck stars can be 
ascribed to the running of the Newton constant. Then, after computing the lower order corrections to the temperature of 
the scale--dependent Hayward solution,
we request that these corrections coincide with that computed using the GUP, which allow us to connect both approaches. 
Specifically, $\beta$ is related with $\epsilon$ by $\beta=(\pi \epsilon /m_{p})^2$. Even more, since $\epsilon\le m_{p}$, we get
$\beta \le \pi^2$. 

Finally, a couple of comments are in order. First, we obtained a GUP parameter which is of order one, as expected from other 
approaches concerning the GUP (see \cite{Elias2017} and references therein). Second, although the numerical
value of $\beta$ is not fixed unambiguosly, in contrast with \cite{Elias2017}, the procedure here employed
allows to connect for the first time the GUP
with scale--dependent gravity, which is expected to describe in an effective way gravitational 
effects beyond the validity of general relativity. Finally, in this sense, and in the context of black holes, we conjecture that 
scale--dependent gravity could encode some features of a generalized 
uncertainty principle, at least in an effective way. Even more, 
we expect that the effect of the running of the couplings 
in the context of scale--dependent gravity may leave some imprint on the Cosmological
Microwave Background in the same way that this effect is expected 
in the GUP scenario \cite{cai2007, cai2014}

\section*{ACKNOWLEDGEMENTS}
The author P.B. was supported by the Faculty of Science and Vicerrector\'{\i}a de Investigaciones 
of Universidad de los Andes, Bogot\'a, Colombia. P. B. dedicates this work to Ana\'{\i}s Dorta--Urra and to Luc\'{\i}a and
In\'es Bargue\~no--Dorta.

\end{document}